# Dissecting coupled orders in a terahertz-driven electron-doped cuprate

Liwen Feng[1,2,3,4*], Haotian Zhang[1*], Tim Priessnitz[2*], Jiayuan Cao[1], Tarapada Sarkar[5], Thales de Oliveira[6], Alexey N. Ponomaryov[6], Igor Ilyakov[6], Fei Yang[7], Yongbo Lv[1], Yuheng Guo[1], Kilian Srowik[3], Steffen Danzenbächer[3], Moritz Niethammer[3], Sergey Kovalev[6], Jan-Christoph Deinert[6], Stefan Kaiser[2,3,4], Richard L. Greene[5], Hao Chu[1†]

[1]*School of Physics and Astronomy, Shanghai Jiao Tong University, Shanghai 200240, China*
[2]*Max Planck Institute for Solid State Research, 70569 Stuttgart, Germany*
[3]*Institute of Solid State and Materials Physics, Dresden University of Technology, 01062 Dresden, Germany*
[4]*4th Physics Institute, University of Stuttgart, 70569 Stuttgart, Germany*
[5]*Maryland Quantum Materials Center, Department of Physics, University of Maryland, College Park, MD 20742 USA*
[6]*Helmholtz-Zentrum Dresden-Rossendorf, Bautzner Landstr. 400, 01328 Dresden, Germany*
[7]*The Pennsylvania State University, University Park, PA, 16802, USA*

*contributed equally to this work
†corresponding author: haochusjtu@sjtu.edu.cn

**Abstract**

The interplay between superconductivity and charge density wave has often been studied from an equilibrium point of view. For example, using static tuning knobs such as doping, magnetic field and pressure, superconductivity can be enhanced or suppressed. The resulting effect on the co-existing charge density wave order, if any, is judged by variations in its ground state properties such as the ordering temperature or the spatial correlation. Such an approach can be understood as coordinated static displacements of two coupled order parameters within a Ginzburg-Landau description, evincing their interplay as either co-operative or competing but does not provide further microscopic information about the interaction. In order to assess such information, we dynamically perturb both orders from equilibrium and observe their coupling directly in the time-domain. We show that high-field multicycle terahertz pulses drive both the Higgs amplitude fluctuations of the superconducting order as well as collective fluctuations of the charge order in an electron-doped cuprate, resulting in characteristic third harmonic generation. A notable time delay is manifested between their respective driven dynamics. We propose that this may signify the important energy scale describing their coupling or imply a terahertz field-depinned charge density wave that destroys macroscopic superconductivity. Our work demonstrates a holistic approach for investigating coupled superconducting and charge density wave orders, which may shed novel light on their intertwined presence and widespread fluctuations in many classes of unconventional superconductors.

**Main**

In materials exhibiting strong electron-phonon coupling, superconductivity and charge density wave (CDW) are often found as alternative ground states of the system. In fact, the dynamics of the latter has also been considered as a possible mechanism behind dissipationless charge transport[1]. Although proven otherwise, such close relationship between the two states continues to be reflected in a wide range of materials beyond mean-field description, including high-temperature superconductors[2], organic superconductors[3], kagome materials[4], discovered in the past several decades. In many of such systems, the termination of a commensurately-ordered and an incommensurately-ordered CDW at a quantum critical point with enhanced quantum fluctuations is found to coincide with the emergence of a superconducting dome[5,6]. This has led many to believe that CDW fluctuations mediate Cooper pairing in such systems[7,8]. Such a perspective highlights the important implications of the dynamical interaction between CDW and superconductivity, apart from their equilibrium manifestations (i.e. whether independent, competing or co-operative). In cuprate high-temperature superconductors, both the superconducting coherence length and the charge order correlation length are remarkably short, thus providing a fertile ground for their collective fluctuations. Therefore, a thorough knowledge about such fluctuations and their dynamical interplay is indispensable for understanding the equilibrium superconducting and CDW states.

Building upon conventional spectroscopy techniques, ultrafast spectroscopy has recently emerged as a novel method to dynamically perturb these orders away from equilibrium and visualize their respective relaxation processes[9-15]. These investigations typically employ a near-infrared pump pulse of ~ 1.5 eV photon energy. The pump pulse excites high-energy electron-hole pairs, often of interband nature, whose subsequent relaxation generates an avalanche of incoherent bosons (phonons, paramagnons, etc.) and low-energy quasiparticles that in turn induce pair-breaking or phase coherence between Cooper pairs. In systems exhibiting a quasi-long-range ordered CDW, a short laser pulse can also impulsively launch CDW amplitude oscillation, with a frequency falling in the terahertz range[10,12]. Information about the dynamical interaction between these order parameters, however, is difficult to obtain due to the lack of simultaneous spectroscopic coupling to those order parameters, which are further obscured by the avalanche of incoherent excitations in other degrees of freedom (e.g. phonons, magnons, etc.).

Recently it was shown that a terahertz pulse, with an energy scale directly relevant for superconductivity and CDW, may circumvent these problems by directly coupling to the collective fluctuations of these two orders[16-24]. In particular, a multicycle terahertz pulse was shown to coherently drive the Higgs amplitude fluctuations of the superconducting order parameter via a two-photon process[17,19,20]. The same terahertz pulse, of frequency $\omega$, can then undergo anti-Stokes scattering with the driven Higgs oscillation of frequency $2\omega$ to generate a third harmonic photon at $3\omega$, which is detected as the spectroscopic signature of the driven collective mode oscillation[17,19,21,22,24]. *Microscopically distinct from mid-/near-IR high harmonic generation in semiconductors, this inelastic scattering process describes in fact to a non-resonant Raman*

*process under the guise of third harmonic generation (THG)*[25-27]. In this case, the same terahertz pulse acts both as the pump and as the probe, ensuring that the excitation and probing processes selectively address the low-energy degrees of freedom of the condensed phases. Indeed, a remarkable sensitivity to the superconducting transition[17], possible above-$T_c$ superconducting fluctuations[19,20], and potential interplays between the Higgs mode and other collective modes[19,21,22,24] has been demonstrated with this technique in various types of superconductors. In this study, we apply such a technique to investigate the Higgs amplitude fluctuations of an electron-doped cuprate $La_{2-x}Ce_xCuO_4$ (LCCO) ($x$ = 0.11, $T_c$ = 27 K), and report evidences for their dynamical interaction with the charge order fluctuations also present in the system. Our experiment is conducted at the TELBE beamline at Helmholtz Center Dresden Rossendorf, where frequency-tunable carrier-envelope-phase-stable multi-cycle terahertz pulses are generated from a superradiant high-field THz source. The terahertz transmission through the sample, consisting of both the driving field and the nonlinear response of the sample, is measured using electro-optical sampling yielding the time profile of the terahertz electric field. The high repetition rate (50 kHz) of the set-up together with an efficient pulse-to-pulse jitter-correction infrastructure allows us to obtain high-quality nonlinear signals with a low driving field strength, enabling our unique observation. A schematic of our optical set-up is shown in the Supplementary Information.

**Temperature and Driving Field Dependence**

We first apply 0.3 THz periodic drive with a peak electric field of 15 kV/cm (corresponding to a power density of ~ 1 µJ/cm² on the sample surface). The THG response of the sample is visualized by applying Fourier bandpass filters to the as-measured terahertz waveforms, which are otherwise dominated by linear transmission (Fig. 1(b)). As Fig. 1(a) shows, a prominent THG response, centered around 25 ps time delay, arises below $T_c$ and grows monotonically as temperature decreases. From the Fourier amplitude spectrum of the raw terahertz transmission, we can extract the amplitude of the THG response for each temperature. Fig. 1(c) shows that the THG amplitude increases steeply below $T_c$ and saturates below 20 K, displaying an order parameter-like temperature dependence consistent with an underlying Higgs fluctuation-mediated THG scattering process (see Supplementary Information for microscopic pictures). The observed temperature dependence implies that the superconducting transition temperature remains unchanged from the equilibrium $T_c$, indicating that the sample stays in the weakly perturbed regime. Above $T_c$, a finite

THG response persists and remains centered around 25 ps time delay. The microscopic origin of it will be further investigated below.

Next, we increase the driving field strength to 41 kV/cm. Compared to the results above obtained in the perturbative regime, with a stronger driving field the time-domain THG response now exhibits an envelope that deviates visibly from a strict Gaussian form. At temperatures like 14 K and 18 K, the THG envelope displays a remarkably flat and elongated top. Such a response can be modeled as the coherent sum of two time-separated Gaussian wavelets with similar amplitudes and commensurate carrier phases. The presence of these two individual wavelets in the full THG response is more clearly revealed at 21.3 K. A noticeable dip can be discerned in the THG envelope around 23 ps time delay, which results from both wavelets decreasing in amplitude as a function of temperature. With further increases in temperature, the earlier-in-time THG response disappears above $T_c$ (we will identify this as the Higgs fluctuation-mediated response). In comparison, the latter-in-time THG response persists to much higher temperatures. By linearly extrapolating the THG amplitude over temperature, we estimate that the latter-in-time THG response vanishes between 200 K and 300 K (we will identify this as the CDW fluctuation-mediated response), which coincides with the onset temperature for the charge order in optimally electron-doped cuprates[28].

**Magnetic Field and Driving Frequency Dependence**

To confirm the microscopic origin of the both THG responses, we place our sample inside a *c*-axis magnetic field and compare the nonlinear response with and without an external magnetic field. Previous studies have shown that a field strength of 7 T almost fully quenches superconductivity in electron-doped cuprates with a similar $T_c$[29]. Indeed, as Fig. 2(a) shows, the overall THG response is quickly suppressed by the magnetic field. At 5 T, the THG waveform exhibits a similar dip in the middle due to the rapid suppression of the Higgs fluctuation-mediated THG response (and the relative insensitivity of the latter-in-time THG response to magnetic field). At 7 T, the earlier-in-time THG response visually vanishes while the latter-in-time THG response persists. If we then measure the temperature dependence of the latter-in-time THG response inside the 7 T magnetic field (Fig. 2(b)), we observe a continuous evolution of the latter-in-time THG response across $T_c$ all the way down to 5 K, exhibiting also an order parameter-like temperature dependence that onsets at a much higher temperature. In addition, the THG response above $T_c$ exhibits very similar evolutions with and without an external magnetic field (Fig. 2(c)). This indicates that the microscopic mechanism responsible for the high-field THG response (i.e. when superconductivity is fully quenched) continues to generate THG above $T_c$ and that the microscopic degree of freedom behind is little affected by an external magnetic field. Such behavior is indeed reminiscent of the weak magnetic field dependence of the charge order in electron-doped cuprates as reported by x-ray scattering investigations[28].

To gain further insights, we also vary the terahertz driving frequency. Fig. 3 shows the sample's nonlinear response to 0.5 THz and 0.7 THz periodic drives with a similar peak electric field around 50 kV/cm. Consistent across all three driving frequencies, the earlier-in-time THG response onsets below $T_c$ and precedes the latter by approximately 5 ps. Interestingly, whereas the two THG responses interfere constructively under 0.3 THz drive, their interference evolves into destructive when the driving frequency is changed towards 0.7 THz. Such a destructive interference in the time domain manifests also as a splitting of the THG peak in the Fourier spectrum (Fig. 3(b)(e)). The amplitude of the full nonlinear response (Fig. 3(c)(f)), obtained by integrating the total area under the THG peak, likewise manifests two distinct temperature-dependent contributions: one originating right at $T_c$ and the other originating above 200 K, corroborating the results obtained with 0.3 THz drive.

Since the THG response under 0.7 THz periodic drive displays two easily-visualized wavelets as a result of their destructive interference, it allows a more careful separation of the magnetic field dependence of each THG response. As Fig. 4(a)(b) show, with a magnetic field greater than 3 T applied along the $c$-axis of the sample, the earlier-in-time THG response becomes barely visible in both the time and frequency domains. By setting two fixed time-windows for separately integrating the earlier-in-time THG amplitude and the latter-in-time THG amplitude, we find that the earlier-in-time THG response becomes monotonically suppressed while the latter-in-time THG response becomes slightly enhanced by turning on the magnetic field (Fig. 4(c)). In particular, beyond 1 T magnetic field the latter exhibits a relatively weak change (less than 15 % increase in amplitude) in contrast to the Higgs-mediated earlier-in-time THG response (nearly 100% reduction). This relatively weak magnetic field dependence of the latter-in-time THG response again corroborates with the results obtained above $T_c$ under 0.3 THz drive.

**Discussions**

**Identifying the microscopic mechanisms behind the two THG responses**

Previous phase-resolved nonlinear terahertz investigations of hole-doped cuprate superconductors have uncovered a unique evolution of the THG response below $T_c$ manifesting a characteristic Fano interference or level repulsion[19,21,24], both indicative of the presence of two independent degrees of freedom partaking in the inelastic scattering process, in stark contrast to *s*-wave superconductors which manifest only a single THG response. In these works, a range of collective modes have been considered for the terahertz- or meV-energy scale scattering process inside cuprates, including the superconducting Higgs fluctuations, charge order fluctuations, Josephson phase fluctuations, a putative pseudogap collective mode, as well as paramagnons and the Bardasis Schrieffer mode. Identifying the collective modes involved in the THG process could be important for unveiling the mechanisms behind high-temperature superconductivity as well as the disparate phases accompanying the superconducting state.

In presenting our experimental results above, we have corroborated the temperature and magnetic field dependences of the two THG responses with previous knowledge about the charge order and superconductivity in electron-doped cuprates. In our previous investigation of $2H$-NbSe$_2$, we have shown that CDW fluctuations may mediate terahertz THG in a similar way as Higgs amplitude fluctuations, as theory also predicts[22,27]. Phenomenologically, in microwave transport studies, CDW is found to give rise to strongly nonlinear AC conductivity, characterized by Shapiro steps, high- and sub-harmonics generation above an electrical field threshold[30]. Our previous study of hole-underdoped La$_{2-x}$Sr$_x$CuO$_4$, where the CDW is quasi-long-range-ordered, similarly reveals two distinct THG responses[31]. In consideration of all these facts, we ascribe the latter-in-time THG response to the charge order fluctuation-mediated THG process. A third mechanism for terahertz THG, involving purely single-particle density fluctuations, has also been experimentally reported for semiconductors[22,32] (see Supplementary Information for details and illustration). Such a process depends sensitively on the carrier density (including thermally-excited carriers) around the Fermi surface and is expected to be resonantly enhanced by specific features in their joint density of states (for example band nesting). In the case of superconductors, it was pointed out that THG can be resonantly enhanced when terahertz two-photon absorption ($2\omega$) coincides with the particle-hole excitation across the superconducting gap ($2\Delta$)[33]. We can rule out this single-particle mechanism as contributing to the latter-in-time THG response since the latter evolves continuously across the superconducting transition without exhibiting any anomaly that may be associated with the gap-opening on the Fermi surface.

We also briefly discuss the origin of the earlier-in-time THG response which onsets in the perturbative driven limit. Such a perturbative THG response has been thoroughly characterized in previous studies on similar superconducting systems. A nearly isotropic response in the terahertz THG channel[19] and a similar $\chi^{(3)}(\omega_{IR}+2\omega_{THz}, \omega_{IR}, \omega_{THz}, \omega_{THz})$ channel[20] was reported for hole-doped cuprates below $T_c$, consistent with a dominant $A_{1g}$ Higgs mode contribution to the third-order nonlinear signal. Theoretically, it has been suggested that in clean superconductors befitting a mean field description, single-particle density fluctuation contributes most THG, while for dirty and correlated superconductors the Higgs fluctuation-mediated contribution prevails[34-36]. We note that electron-doped cuprates are generally considered as dirty superconductors[37,38] and as such, a Higgs fluctuation-mediated earlier-in-time THG process would also be consistent with theory. In addition, a $\chi^{(3)}(\omega_{THz}, -\omega_{THz}, \omega_{THz}, \omega_{THz})$ contribution to the fundamental response accompanying the $\chi^{(3)}(3\omega_{THz}, \omega_{THz}, \omega_{THz}, \omega_{THz})$ THG response can also be inferred from our results (see Supplementary Information). Such a $\chi^{(3)}$ nonlinear fundamental response has been recently proposed as a more unambiguous signature of the Higgs mode[39].

**Time delay between the two THG responses**

We turn our attention to the appearance of the latter-in-time THG response at high driving field strength. First, we note that the appearance of a second dynamical process (i.e. in addition to the dynamics of the Cooper pairs condensate) with increasing pump fluences has been similarly noted in several earlier optical pump-probe experiments[11,40]. For example, in Ref. 11, the transient reflectivity of a near-optimally doped $Nd_{2-x}Ce_xCuO_{4+\delta}$ after photo-excitation by 1.5 eV pump pulses of 2 $\mu J/cm^2$ fluence displays a positive change ($\Delta R/R$) and a slow relaxation dynamics that is attributed to the recombination of broken Cooper pairs. With stronger pump fluences like 4 $\mu J/cm^2$ and 6 $\mu J/cm^2$, a negative component in $\Delta R/R$ becomes increasingly visible, exhibiting a faster relaxation and is attributed to Raman-like collective fluctuations of a competing order. Since both the pump and the probe photon energy resides far above the characteristic energy scale of these collective modes, the optical relaxation $\Delta R(t)/R$ could be dominated by incoherent processes in diverse degrees of freedom. As such, the dynamical interaction between the two orders is not coherently visualized.

In comparison, the low-energy terahertz field employed in our study directly and coherently addresses the collective dynamics of the two orders, thereby enabling a clear observation of the time delay separating their driven responses. Borrowing insights from a simple driven coupled oscillators model, equivalent to a dynamical Ginzbug-Landau model with coupled order parameters[22], we can see that the time delay between the two oscillators' responses depends sensitively on the coupling strength (Fig. 4(d)). In particular, a weak coupling results in a slower energy transfer and a longer delay between the two oscillators' responses, while a strong coupling leads to almost simultaneous responses. Within this interpretation, the visual appearance of the charge order fluctuation-mediated THG response at larger terahertz field strength can be rationalized as due to a weak coupling between the two orders, consistent with conventional wisdom about electron-doped cuprates. That is, a strong driving field has to be employed in order for the charge order fluctuation to build up enough amplitude to be spectroscopically resolved. Along this interpretation, a possibility worth investigating is whether the interaction contains non-biquadratic terms (e.g. $g\Phi^2\Psi$ in the Ginzbug-Landau description) leading to nonlinearly coupled equations of motion[22]. This would imply that the amplitude of the charge order fluctuation gets nonlinearly enhanced as the amplitude of the Higgs fluctuation increases. Finally, we cannot exclude another distinct scenario: above a certain electrical field threshold, the terahertz pulse de-pins the charge order from local defects and results in a sliding CDW[30]. Within this interpretation, the fact that the Higgs-mediate THG response disappears as soon as the charge order fluctuation-mediated THG process emerges implies that a sliding CDW and macroscopic superconductivity are mutually exclusive[14,15], possibly due to the destruction of the superfluid phase coherence by the sliding CDW. We note that in hole-optimally-doped cuprates and conventional superconductors where a long-range ordered CDW is absent, a well-defined Higgs fluctuation-mediated THG response persists up to 500 kV/cm peak electrical field[19,20,23], in stark contrast to systems where the two orders co-exist. This is also in-line with the view that the superconducting phase coherence is weak in hole-underdoped cuprates where a quasi-long-range-ordered CDW is present.

In conclusion, using terahertz third harmonic generation, a non-resonant Raman scattering technique under guise, we observe two distinct THG responses from an optimally-doped LCCO thin film, arising respectively from the coherently-driven Higgs and charge order fluctuations. While previous optical pump probe studies have similarly reported a two-component relaxation process in electron-doped cuprate, ascribed respectively to the Cooper pair recombination and the relaxation of a competing order, the terahertz THG responses reported in our study exhibit characteristic temperature- and magnetic field-dependences that closely identify with the superconducting order and charge order in electron-doped cuprates. In addition, for the first time a time delay between such two dynamics is visualized, providing a unique microscope into the microscopic scattering process that mediates the interaction between the two collective modes. We propose that such a time delay signifies important information regarding the interaction energy scale. Such information might be essential for understanding many peculiar aspects of the cuprate phase diagram including the microscopic nature of the pseudogap[41], a possible pair density wave state underlying the stripe order[42], the origin of ubiquitous superconducting and charge order fluctuations above the superconducting dome[43,44], etc. With these rich prospects and the unique capabilities of nonlinear terahertz spectroscopy as demonstrated above, we hope to stimulate future applications of this technique to a wider class of materials with more kinds of intertwined orders.

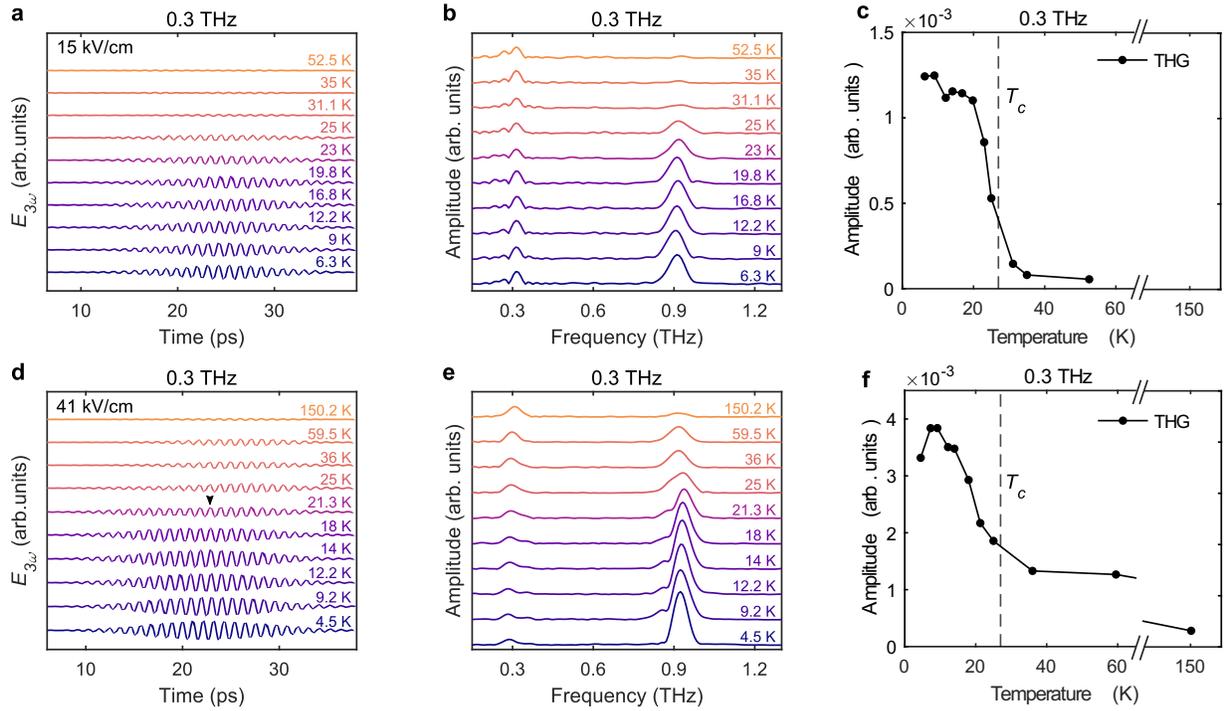

**Figure 1 THG response of LCCO (x = 0.11, $T_c$ = 27 K) under perturbative and non-perturbative 0.3 THz drive (a)** Time-domain THG waveforms extracted from the raw terahertz transmission using Fourier bandpass filters with a peak incidence field strength of 15 kV/cm. A pronounced THG response emerges below $T_c$, increasing monotonically as the temperature decreases. **(b)** Fourier amplitude spectra for the raw transmission data behind (a). **(c)** Temperature dependence of the THG amplitude extracted from the Fourier spectra in (b). **(d)-(f)** Corresponding results obtained with 41 kV/cm peak terahertz field. In (d), the presence of two time-delayed THG responses is revealed by the waveform at 21.3 K, where a dip (marked by triangle) in the envelope function of the full oscillatory response is manifested. At lower temperatures, the two time-delayed oscillatory responses extend further in time, producing an elongated envelope function for example at 18 K.

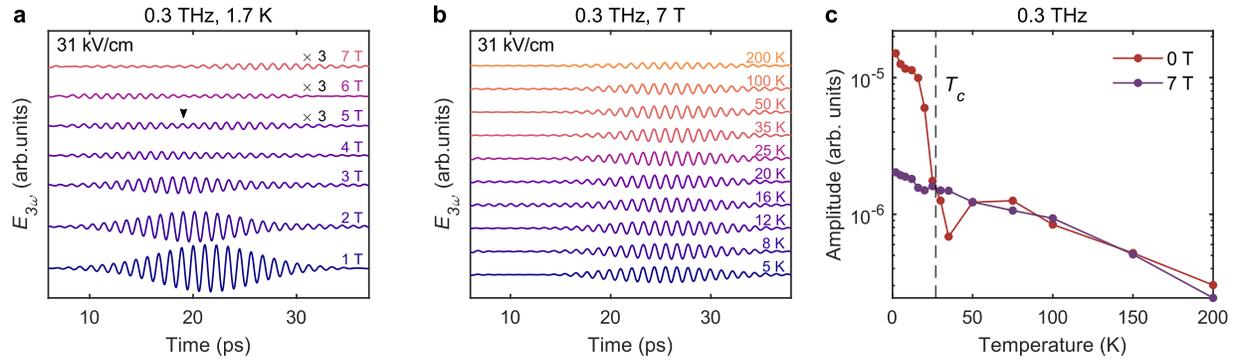

**Figure 2 Magnetic field dependence of THG response under 0.3 THz drive (a)** Time-domain THG response to a *c*-axis-oriented magnetic field under 0.3 THz periodic drive. The overall THG response is suppressed by the magnetic field. At 5 T, the THG waveform visually displays two time-delayed components. A triangle marks the dip in their middle. At 7 T, only the latter-in-time THG response remains. **(b)** Time-domain THG response under non-perturbative 0.3 THz periodic drive inside a 7 T magnetic field. **(c)** Temperature dependence of the Fourier amplitude of the total THG response, in the presence and absence of a 7 T field.

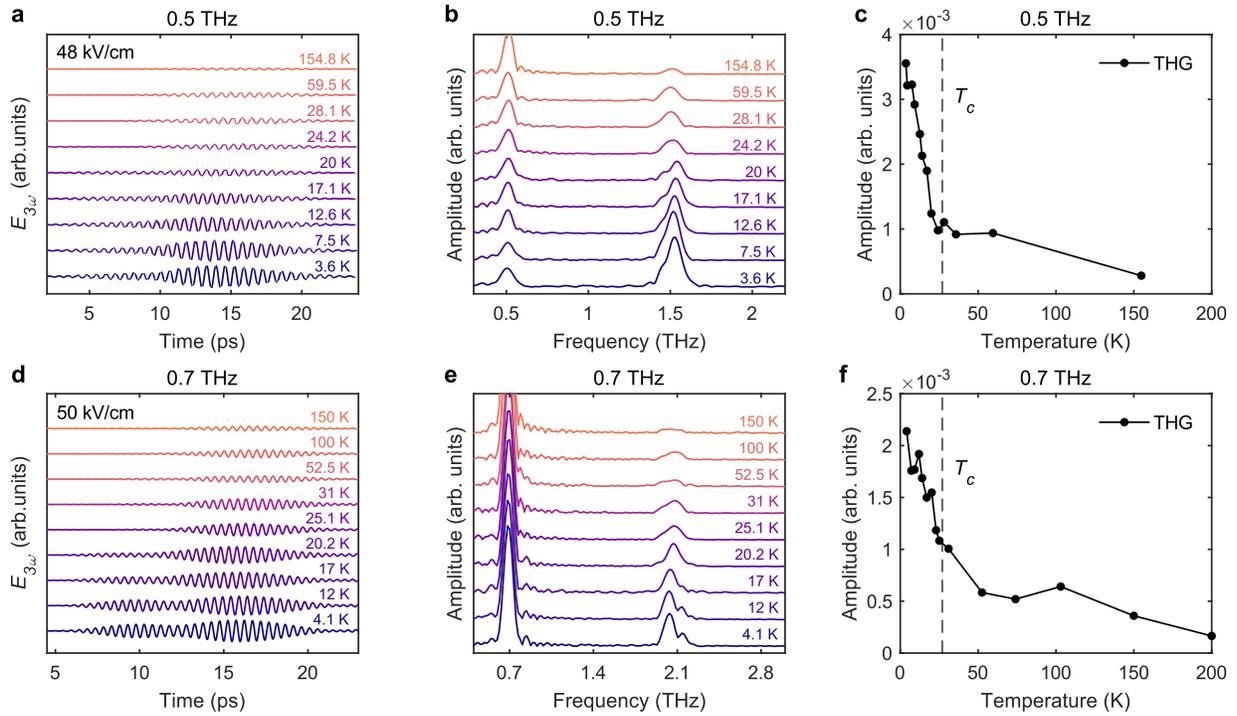

**Figure 3 THG response of LCCO under non-perturbative 0.5 THz and 0.7 THz drive (a)** Time-domain THG response obtained with 0.5 THz periodic drive with a peak incident field strength of 48 kV/cm. The presence of two time-delayed THG responses becomes more visible here compared to Fig. 1(d). **(b)** Fourier amplitude spectrum of the terahertz transmission and **(c)** the extracted THG amplitude as a function of temperature. **(d)-(f)** Corresponding results obtained with 0.7 THz periodic drive and a peak incident field strength of 50 kV/cm. In (d) the destructive interference between the two time-delayed THG responses makes their presence very visible.

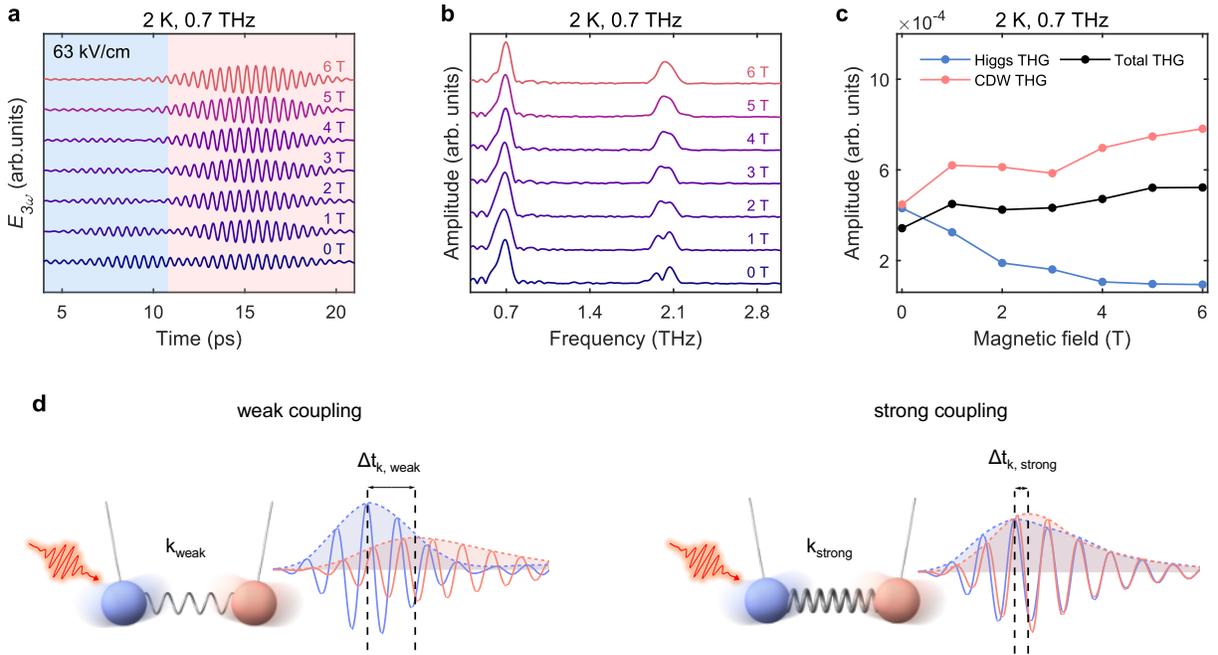

**Figure 4 Magnetic field dependence of the THG response** **(a)** Time-domain THG response of the sample to a *c*-axis magnetic field under 0.7 THz periodic drive. The magnetic field suppresses the earlier-in-time THG response while just slightly enhancing the latter-in-time THG response. **(b)** Corresponding Fourier amplitude spectra for the raw transmission data behind (a). **(c)** The total THG amplitude, earlier-in-time THG amplitude, and later-in-time THG amplitude extracted from their respective time-windows (the two shaded areas) in (a) as a function of magnetic field. **(d)** Dynamical responses of two coupled oscillators. The blue pendulum is directly driven by a Gaussian-enveloped periodic drive. The red pendulum is indirectly driven via its coupling, a spring, to the blue pendulum. Left: if the coupling is weak, the red pendulum reaches maximum displacement at longer time delay with respect to the blue pendulum. Right: if we increase the coupling constant, the two pendula exhibit more similar driven dynamics. In particular, the time delay between reaching their maximum displacements decreases towards 0.

**Acknowledgments**

The authors thank the ELBE team for the operation of the TELBE facility and Sida Tian for valuable discussions. H. C. acknowledges support by the National Key Research and Development Program of China (Grants No. 2024YFA1408701), the National Natural Science Foundation of China (Grants No. 12274286) and the Yangyang Development Fund. R. L. G. and T. S. acknowledge support by NSF grant (# DMR 2002658). S. Ka. acknowledges funding by the European Union (ERC, T-Higgs, Grant No. GA 101044657), Deutsche Forschungsgemeinschaft (DFG) through SFB 1143 (project id 247310070); the Würzburg- Dresden Cluster of Excellence on Complexity and Topology in Quantum Matter – ct.qmat (EXC 2147, project id 390858490). Views and opinions expressed are however those of the author(s) only and do not necessarily reflect those of the European Union or the European Research Council Executive Agency. Neither the European Union nor the granting authority can be held responsible for them.

**Author Contributions**

H. C. conceived the experiment. T. S. and R. L. G. synthesized and characterized the sample. L. F., T. P., H. Z., J. C., Y. L., S. Ka. and H. C. conducted the beamtime experiment which is operated by J.-C. D., S. Ko., T. O., A. N. P. and I. I.. L. F. and J. C. performed data analysis. H. Z. and L. F. performed theoretical modeling. H. C., L. F., S. Ka. and F. Y. interpreted the results. H.C. and L. F. wrote the manuscript with H. Z., T. P. and S. Ka., with input from all authors.

**Competing interests**

The authors declare no competing interests.